\documentclass[apjl]{emulateapj}

\usepackage{natbib}
\usepackage{amsmath}
\usepackage{amssymb}
\usepackage[bf, sf, FIGTOPCAP, nooneline, tight]{subfigure}
\usepackage{graphicx}
\usepackage{dcolumn}
\usepackage{bm}
\usepackage[usenames,dvipsnames]{color}
\usepackage[colorlinks=true, linkcolor=BrickRed, citecolor=Blue, urlcolor=Blue, filecolor=Blue]{hyperref}
\usepackage{booktabs}
\usepackage{hhline}
\usepackage{epsf}
\usepackage{color}
\usepackage{epsf}
\usepackage{psfrag}
\usepackage{epsfig}

\usepackage{multirow}
\usepackage{bm}
\usepackage{latexsym}
\usepackage{ifsym}
\usepackage{tabularx}

\usepackage{threeparttable}

\newcommand{\be}{\begin{equation}}
\newcommand{\ee}{\end{equation}}
\def\beq{\begin{equation}}
\def\eeq{\end{equation}}
\def\ber{\begin{eqnarray}}
\def\eer{\end{eqnarray}}

\def \lleq {\lower0.9ex\hbox{ $\buildrel < \over \sim$} ~}
\def \ggeq {\lower0.9ex\hbox{ $\buildrel > \over \sim$} ~}

\def\deg{\ifmmode^\circ\else$^\circ$\fi}

\newcommand{\rma}{\rho_m}
\newcommand{\rr}{\rho_r}
\newcommand{\rL}{\rho_\Lambda}
\newcommand{\dG}{\dot{G}}

\newcommand{\drL}{\dot{\rho}_\Lambda}
\newcommand{\dH}{\dot{H}}
\newcommand{\CC}{\Lambda}

\newcommand{\Om}{\Omega_m}
\newcommand{\Omo}{\Omega_{m}}

\newcommand{\ORo}{\Omega_{r}}
\newcommand{\OL}{\Omega_{\Lambda}}

\newcommand{\OLo}{\Omega_{\Lambda}}

\newcommand{\OD}{\Omega_{D}}

\newcommand{\rc}{\rho_c}
\newcommand{\rco}{\rho_{c 0}}

\newcommand{\rmr}{\rho_m}

\newcommand{\rR}{\rho_r}
\newcommand{\rD}{\rho_D}

\newcommand{\rLo}{\rho_{\CC}^0}

\newcommand{\nueff}{\nu_{\rm eff}}
\newcommand{\nueffp}{\nu_{\rm eff}'}

\begin{document}

\title{Hints of dynamical vacuum energy in the expanding Universe}

\author{Joan Sol\`{a}\altaffilmark{1,2}, Adri\`{a} G\'omez-Valent\altaffilmark{1,2}, Javier de Cruz P\'erez\altaffilmark{1}
}

\altaffiltext{1}{High Energy Physics Group, Dept. ECM, Univ. de Barcelona, Av. Diagonal 647, E-08028 Barcelona, Catalonia, Spain}

\altaffiltext{2}{Institut de
Ci\`encies del Cosmos (ICC), Univ. de Barcelona, Av. Diagonal 647, E-08028 Barcelona, Catalonia, Spain}

\email{sola@ecm.ub.edu}
\email{adriagova@ecm.ub.edu}
\email{jdecrupe7@alumnes.ub.edu}






\begin{abstract}
Recently there have been claims on model-independent evidence of
dynamical dark energy. Herein we consider a fairly general class of
cosmological models with a time-evolving cosmological term of the form
$\Lambda(H)=C_0+C_H H^2+C_{\dot{H}} \dot{H}$, where $H$ is
the Hubble rate. These models are well motivated from the
theoretical point of view since they can be related to the general
form of the effective action of quantum field theory in curved
spacetime. Consistency with matter conservation can be achieved by
letting the Newtonian coupling $G$ change very slowly with the
expansion. We solve these dynamical vacuum models and fit them to
the wealth of expansion history and linear structure formation data.
The results of our analysis indicate a significantly better agreement as
compared to the concordance $\CC$CDM model, thus supporting the
possibility of a dynamical cosmic vacuum.

\end{abstract}


\section{Introduction}
The positive evidence that our Universe is speeding up
owing to some form of dark energy (DE) pervading all corners of the
interstellar space seems to be nowadays beyond doubt after the first
measurements of distant supernovae \citep{SNIaRiess,SNIaPerl}  and the
most recent analysis of the precision cosmological data by the
Planck collaboration\,\citep{Planck}. The ultimate origin of such
positive acceleration is unknown but the simplest possibility would
be the presence of a tiny and positive cosmological constant (CC) in Einstein's
field equations, $\CC>0$. This framework, the so-called concordance
or $\Lambda$CDM model, seems to describe quite well the observations
\citep{Planck} but, unfortunately, there is little theoretical
motivation for it. The CC is usually associated to the energy
density carried by the vacuum, through the parameter
$\rL=\CC/(8\pi\,G)$ (in which $G$ is the Newtonian coupling),
although it is difficult to reconcile its measured value ($\rL\sim
10^{-47}$ GeV$^4$) with typical expectations in quantum field theory
(QFT) and string theory, which are many orders of magnitude bigger.
Such situation has triggered in the past -- only to find it reinforced
at present -- the old CC problem and the cosmic coincidence
problem\,\citep{Weinberg,CCP1,CCP2,CCP3}. Both problems lie at the forefront of
fundamental physics.

Different theoretical scenarios have been proposed since long.
In this Letter we take seriously the
recent observational hint that the DE could be dynamical as a means
to alleviate some tensions recently observed with the
$\CC$CDM\,\citep{SahShaSta}.
%
%
Specifically, we focus on the dynamical vacuum models of the form
$\Lambda(H)=C_0+C_H H^2+C_{\dot{H}} \dot{H}$, in which $H=\dot{a}/a$
and $\dot{H}=dH/dt$ are the Hubble rate and its cosmic time
derivative, with $C_0\neq 0$ a constant. We assume that at least
one of the coefficients $C_H$ and $C_{\dot{H}}$ is nonvanishing.
Such models possess a well-defined $\CC$CDM limit ($C_H, C_{\dot{H}}\to 0$) and involve two time derivatives of  the scale
factor, thereby  can be consistent with the general covariance of the
effective action of QFT in curved spacetime. While the general
structure of $\Lambda(H)$ can be conceived  as an educated
phenomenological ansatz, it can actually be related to the quantum effects on the effective vacuum action due to the expanding background, in which
the leading effects may generically be captured from a
renormalization group equation\,\citep{Fossil07,ShapSol09,JSPRev2013, SolGom2015,Sola2015}. The dimensionless coefficients
$C_H$ and $C_{\dot{H}}$ are actually related to the $\beta$-function
of the running and are therefore naturally
small. In the presence of matter conservation this is possible by letting $G=G(H)$ be dynamical as well\,\citep{Fossil07,JSPRev2013}. A generalization of $\rL(H)$
with higher powers of the Hubble rate, i.e. $H^n\,(n>2)$, has
been recently used to describe inflation, see
e.g. \citep{LimBasSol,Sola2015}.

 In the following we solve these dynamical
vacuum models $\CC(H)$ and test them in the light of the recent
observational data, and compare
their performance with the concordance $\CC$CDM model.

\begin{table*}
 \caption{Best-fit values for G1-type models}
\begin{center}
\resizebox{1\textwidth}{!}{
\begin{tabular}{| c  |c | c | c | c | c |c | c | c | c | c |c |}
\multicolumn{1}{c}{Model} & \multicolumn{1}{c}{$|\frac{\Delta G}{G_0}|$ (BBN,CMB), Omh$^2$}  & \multicolumn{1}{c}{$\Omega_m$} & \multicolumn{1}{c}{$\overline{\Omega}_m$} ({\scriptsize all data}) &  \multicolumn{1}{c}{{\small$\nu$}} &  \multicolumn{1}{c}{{\small$\bar{\nu}$}}  & \multicolumn{1}{c}{$\sigma_8$}  & \multicolumn{1}{c}{$\overline{\sigma}_{8}$}  &
\multicolumn{1}{c}{$\chi^2/dof$} &
\multicolumn{1}{c}{$\overline{\chi}^2/dof$} &
\multicolumn{1}{c}{AIC} &
\multicolumn{1}{c}{$\overline{\rm AIC}$}
\\\hline {\small $\CC$CDM} & -, Yes & {\small$0.278^{+0.005}_{-0.004}$} & {\small$0.276\pm 0.004$} & - & - & {\small$0.815$} & {\small$0.815$} & {\small$828.84/1010$} & {\small$828.69/1010$} & {\small$830.84$} & {\small$830.69$}
\\\hline
{\small G1} & (10\%,5\%), Yes  & {\small$0.278\pm 0.006$} & {\small$0.275\pm 0.004$} & {\small$0.0015^{+0.0017}_{-0.0015}$} & {\small$0.0021^{+0.0014}_{-0.0016}$}  & {\small$0.797$} & {\small$0.784$}  & {\small$822.82/1009$} & {\small$821.97/1009$} & {\small$826.82$} & {\small$825.97$}

\\\hline
{\small $\CC$CDM} & -, No & {\small$0.292\pm 0.008$} & {\small$0.286\pm 0.007$} & - & - & {\small$0.815$} & {\small$0.815$} & {\small$583.38/604$} & {\small$582.74/604$} & {\small$585.38$} & {\small$584.74$}
\\\hline
{\small G1} & (10\%,5\%), No  & {\small$0.290\pm 0.011$} & {\small$0.281\pm 0.005$} & {\small$0.0008^{+0.0016}_{-0.0015}$} & {\small$0.0015\pm 0.0014$}  & {\small$0.795$} & {\small$0.771$}  & {\small$577.62/603$} & {\small$575.70/603$} & {\small$581.62$} & {\small$579.70$}
\\\hline {\small $\CC$CDM*} & -, Yes$^*$ & {\small$0.297\pm 0.006$} & {\small$0.293\pm 0.006$} & - & - & {\small$0.815$} & {\small$0.815$} & {\small$806.68/982$} & {\small$806.17/982$} & {\small$808.68$} & {\small$808.17$}
\\\hline
{\small G1*} & (10\%,5\%), Yes$^*$  & {\small$0.296\pm 0.009$} & {\small$0.287\pm 0.004$} & {\small$0.0006\pm 0.0015$} & {\small$0.0012^{+0.0014}_{-0.0013}$}  & {\small$0.803$} & {\small$0.770$}  & {\small$802.66/981$} & {\small$799.15/981$} & {\small$806.66$} & {\small$803.15$}
\\\hline
 \end{tabular}}
\end{center}
\label{tableFit1}
\begin{scriptsize}
\tablecomments{The best-fitting values for the G1-type models and their
statistical  significance ($\chi^2$-test and Akaike information criterion AIC, see the text).
All quantities with a bar involve a fit to the total input data, i.e. the expansion history  (Omh$^2$+BAO+SNIa), CMB shift parameter, the indicated constraints on the value of $\Delta G/G_0$ at BBN and at recombination, as well as the linear growth data. Those without bar correspond to a fit in which we use all data  but exclude the growth data points from the fitting procedure. ``Yes'' or ``No'' indicates if Omh$^2$ enters or not the fit. The starred scenarios correspond to removing the high redshift point $z=2.34$ from Omh$^2$ (see text). The quoted number of degrees of freedom ($dof$) is equal to the number of data points minus the number of independent fitting parameters. The fitting parameter $\nu$ includes all data.}
\end{scriptsize}
\end{table*}

\begin{table*}
 \caption{Best-fit values for G2-type models}
\begin{center}
\resizebox{1\textwidth}{!}{
\begin{tabular}{| c  |c | c | c | c | c |c | c | c | c | c |c |}
\multicolumn{1}{c}{Model} & \multicolumn{1}{c}{$|\frac{\Delta G}{G_0}|$ (CMB), Omh$^2$}  & \multicolumn{1}{c}{$\Omega_m$} & \multicolumn{1}{c}{$\overline{\Omega}_m$} ({\scriptsize all data}) &  \multicolumn{1}{c}{{\small$\nueff$}} &  \multicolumn{1}{c}{{\small$\bar{\nu}_{\rm eff}$}}  & \multicolumn{1}{c}{$\sigma_8$}  & \multicolumn{1}{c}{$\overline{\sigma}_{8}$} &
\multicolumn{1}{c}{$\chi^2/dof$} &
\multicolumn{1}{c}{$\overline{\chi}^2/dof$} &
\multicolumn{1}{c}{AIC} &
\multicolumn{1}{c}{$\overline{\rm AIC}$}
\\\hline {\small $\CC$CDM} & -, Yes & {\small$0.278^{+0.005}_{-0.004}$} & {\small$0.276\pm 0.004$} & - & - & {\small$0.815$} & {\small$0.815$} & {\small$828.84/1009$} & {\small$828.69/1009$} & {\small$830.84$} & {\small$830.69$}
\\\hline
{\small G2} & 5\%, Yes & {\small$0.278\pm 0.006$} & {\small$0.277\pm 0.004$} & {\small$0.0038^{+0.0025}_{-0.0023}$} & {\small$0.0043^{+0.0018}_{-0.0020}$}  & {\small$0.774$} & {\small$0.773$} & {\small$817.17/1008$} & {\small$817.26/1008$} & {\small$821.17$} & {\small$821.26$}
\\\hline
{\small $\CC$CDM} & -, No & {\small$0.292\pm 0.008$} & {\small$0.286\pm 0.007$} & - & - & {\small$0.815$} & {\small$0.815$} & {\small$583.38/603$} & {\small$582.74/603$} & {\small$585.38$} & {\small$584.74$}
\\\hline
{\small G2} & 5\%, No  & {\small$0.287\pm 0.011$} & {\small$0.283\pm 0.005$} & {\small$0.0025^{+0.0026}_{-0.0025}$} & {\small$0.0030^{+0.0021}_{-0.0018}$} & {\small$0.763$} & {\small$0.767$} & {\small$572.68/602$} & {\small$572.99/602$} & {\small$576.68$} & {\small$576.99$}
\\\hline {\small $\CC$CDM*} & -, Yes$^*$ & {\small$0.297\pm 0.006$} & {\small$0.293\pm 0.006$} & - & - & {\small$0.815$} & {\small$0.815$} & {\small$806.68/981$} & {\small$806.17/981$} & {\small$808.68$} & {\small$808.17$}
\\\hline
{\small G2*} & 5\%, Yes$^*$  & {\small$0.295\pm 0.009$} & {\small$0.289\pm 0.005$} & {\small$0.0015^{+0.0026}_{-0.0025}$} & {\small$0.0028^{+0.0018}_{-0.0021}$} & {\small$0.789$} & {\small$0.765$} & {\small$798.85/980$} & {\small$797.05/980$} & {\small$802.85$} & {\small$801.05$}
\\\hline
 \end{tabular}}
\end{center}
\label{tableFit2}
\begin{scriptsize}
\tablecomments{As in Table 1, but for G2 models with $\xi'=1$ so as to maximally preserve the BBN bound (see text). The effective G2-model fitting parameter in this case is  $\nueff=\nu/4$. The constraint on $|\Delta G/G_0|$ from CMB anisotropies at recombination is explicitly indicated.}
\end{scriptsize}
\end{table*}


\section{Background cosmological solutions}\label{sect:Background}
The field equations for the dynamical vacuum energy density in the
Friedmann-Lema\^\i tre-Robertson-Walker (FLRW) metric in flat space are
derived in the standard way and are formally similar to the ones
with strictly constant $G$ and $\CC$ terms:
\begin{eqnarray}\label{eq:FriedmannEq}
&&3H^2=8\pi\,G(H)\,(\rho_m+\rR+\rho_\Lambda(H))\\
&&3H^2+2\dot{H}=-8\pi\,G(H)\,(p_\Lambda+p_r)\,, \label{eq:PressureEq}\,
\end{eqnarray}
where  $\rL(H)={\Lambda(H)}/(8\pi G(H))$ is the dynamical vacuum energy density, $p_\Lambda(H)=-\rho_\Lambda(H)$, and
$G(H)$ is the dynamical gravitational coupling. It is convenient to take into account from the beginning the effect of relativistic
matter, i.e. $p_r=(1/3)\rho_r$,  together with dust ($p_m=0$).  We consider the
following two realizations of the dynamical vacuum model:
\begin{eqnarray}
G1: \phantom{XXX} \Lambda(H)&=&3(c_0+\nu
H^2)\label{eq:G1}\\
G2: \phantom{Xx}\Lambda(H,\dot{H})&=&3(c_0+\nu H^2+\frac{2}{3}\alpha\,\dH)\,,\label{eq:G2}
\end{eqnarray}
where we have redefined $C_0=3c_0$, $C_H= 3\nu$ and $C_{\dot{H}}=2\alpha$ for convenience. Model G1 is of course a particular case of Model G2, but it will be useful to distinguish between them. We can
combine \eqref{eq:FriedmannEq} and \eqref{eq:PressureEq} to obtain
the equation of local covariant conservation of the energy, i.e.
$\nabla^\mu(GT_{\mu 0})=0$. Explicitly, since we assume matter
conservation (meaning $\dot\rho_m+3H\rmr=0$ and $\dot\rho_r+4H\rR=0$), it leads to a
dynamical interplay between the vacuum and the Newtonian coupling:
\be\label{eq:Bianchi}\dG(\rma+\rr+\rL)+G\drL=0. \ee
Trading the cosmic time for the scale factor $a$,
the previous equations amount to determine $G$ as a function of $a$. Using the matter conservation equations, we arrive at
%
\be\label{eq:Star} G(a)=- G_0\,\left[\frac{a\,\left(E^2(a)\right)^{\prime}}{3 \Omo\,a^{-3}+4\ORo\,a^{-4}}\right]\,, \ee
where $G_0\equiv G(a=1)$ is the present value of $G$, and $E(a)=H(a)/H_0$ is the normalized Hubble rate to its present value. The prime stands for $d/da$, and $\Omega_{i}=\rho_{i 0}/\rco$  (with $\rco=3H_0^2/8\pi G_0$) are the currently normalized energy densities with respect to the critical density. Inserting (\ref{eq:G2}) and the above result for $G(a)$ in Eq.\,(\ref{eq:FriedmannEq}) and integrating, we obtain:
\be\label{eq:DifEqH} E^2(a)=1+\frac{\Omo}{\xi}\left[-1+a^{-4\xi'}\left(a+\frac{\xi\,\ORo}{\xi'\Omo}\right)^{\frac{\xi'}{1-\alpha}}\right]\,, \ee
where have introduced
\be\label{eq:xixip}
\xi=\frac{1-\nu}{1-\alpha}\equiv 1-\nueff\,,\ \ \  \xi^\prime=\frac{1-\nu}{1-\frac{4}{3}\alpha}\equiv 1-\nueffp\,.
\ee
For small $|\nu,\alpha|\ll1$ (the expected situation), we can use the approximations $\nueff\simeq \nu-\alpha$ and $\nueffp\simeq \nu-(4/3)\alpha$. Note that, in order to simplify the presentation, we have removed from Eq.\,(\ref{eq:DifEqH}) terms proportional to $\ORo\ll\Omo$ that are not relevant here. We can check e.g. that in the radiation-dominated epoch the leading term in the expression (\ref{eq:DifEqH}) is $\sim \ORo\,a^{-4\xi'}$, whilst in the matter-dominated epoch is $\sim \Omo\,a^{-3\xi}$. Furthermore, we find that the (full) expression for $E^2(a)$  reduces to the $\CC$CDM form, $1+\Omo\,(a^{-3}-1)+\ORo(a^{-4}-1)$, in the limit $\nu,\alpha\to 0$ (i.e. $\xi,\xi'\to 1$).
Notice also the constraint among the parameters,
$
c_0=H_0^2\left[\OLo-\nu+\alpha\left(\Omo+\frac43\,\ORo\right)\right]\,,
$
which follows
from matching the vacuum energy density $\rL(H)$ to its present
value $\rLo$ for $H=H_0$ and using $\Omo+\ORo+\OLo=1$.  The explicit scale factor dependence of the Newtonian coupling
ensues upon inserting  \eqref{eq:DifEqH} in \eqref{eq:Star} and computing the derivative. We refrain once more from writing out the full expression here, but one can check that in the limit $a\to 0$ (relevant for the Big Bang Nucleosynthesis epoch) it behaves as
\begin{figure*}
\centering
\includegraphics[angle=0,width=0.5\linewidth]{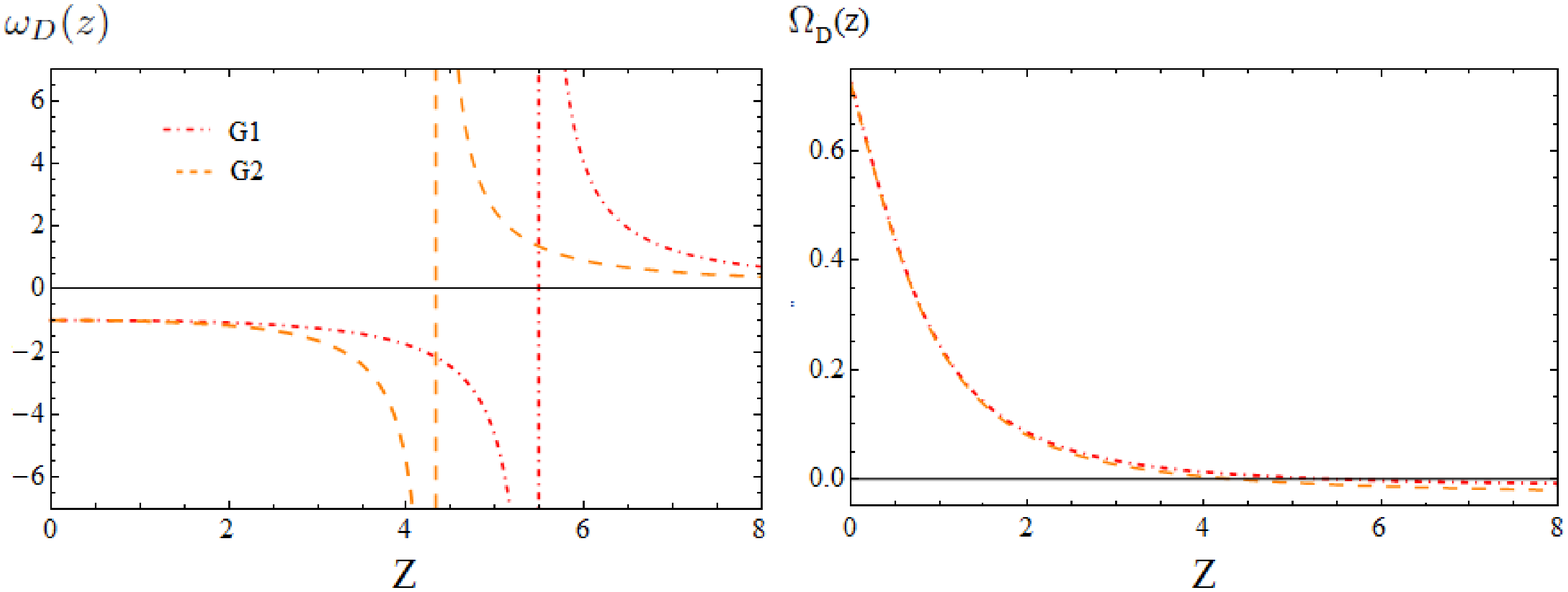}
\caption{\label{fig:DE}%
\scriptsize Left: Evolution of the effective EoS $\omega_D(z)$, Eq.\,(\ref{eq:effEoS}), for the models under consideration: Right: The corresponding evolution of the effective DE density $\OD(z)$ normalized to the critical density (see text).
\vspace{0.3cm}
}
\end{figure*}

\be\label{eq:GNR} G(a)=G_0\,a^{4(1-\xi')}\simeq G_{0}(1+4\nueffp\,\ln\,a)\,. \ee
Thus, the gravitational coupling
evolves logarithmically with the scale factor and hence changes very slowly. This
logarithmic law was motivated previously in \citep{Fossil07,JSPRev2013} within the context of the renormalization group of QFT in curved spacetime.
For $\nu=\alpha=0$  we obtain
$G=G_0$ identically, i.e. the current value of the gravitational coupling. However the situation $G=G_0$ is also attained in the limit $a\to0$ for $\nu=(4/3)\alpha$ (i.e. $\xi'=1)$; and indeed we shall adopt this setting hereafter in order to maximally preserve the BBN constraint for the G2 model. The effective fitting parameter will be $\nueff=\nu/4$.  Obviously this setting is impossible for G1, so in this case we will adopt the average  BBN restriction $|\Delta G/G|<10\%$ in the literature\,\citep{Chiba,Uzan}. At the same time we require  $|\Delta G/G|<5\%$ at recombination ($z\simeq 1100$) for both G1 and G2 from the CMB anisotropy spectrum\,\citep{Chiba}.

The expression for the dynamical vacuum energy density can be
obtained from Friedmann's equation (\ref{eq:FriedmannEq}), in combination with the explicit form of $G(a)$. We quote here only the simplified expression valid for the matter-dominated epoch:
\be\label{eq:RhoLNR} \rL(a)=\rho_{c 0}\,
a^{-3}\left[a^{3\xi}+\frac{\Omega_m}{\xi}(1-\xi-a^{3\xi})\right]\,.
\ee
For $\xi\to 1$ we have $\rL\to\rho_{c 0}(1-\Om)=\rco\OL$  and
we retrieve the $\CC$CDM case with strictly constant
$\rL$. The form (\ref{eq:RhoLNR}) is sufficient to obtain an effective DE density $\rD(z)$ and effective equation of state (EoS) for the DE at fixed $G=G_0$, as conventionally used in different places of the literature -- see e.g. \,\citep{SolaStefancic,Shafieloo2006,Mirage}. We find:
\begin{equation}\label{eq:effEoS}
\omega_D(a)=-\frac{1}{1+\frac{\rmr(a)}{\rL(a)}\frac{G(a)-G_0}{G(a)}}\,.
\end{equation}
In Fig. 1 (left) we plot $\omega_D$ as a function of the cosmic redshift $z=-1+1/a$ for models G1 and G2.
Near our time, $\omega_D$ stays very close to $-1$ (compatible with the $\CC$CDM), but at high $z$ it departs. In the same Fig. 1 (right) we plot $\OD(z)=\rD(z)/\rc(z)$, i.e. the normalized DE density with respect to the critical density  at constant $G_0$.  The asymptotes of $\omega_D$ for each model at  $z>4$ are due to the vanishing of $\OD(z)$ at the corresponding point (as clearly seen in the figure)-- confer the aforementioned references for similar features.


\section{Fitting the models to the observational data}\label{sect:Fit}

Let us now test these models versus observation. First of all, we
use the available measurements of the Hubble function as collected in \citep{Ding}. These are essentially the data points of \citep{Farooq} in the redshift range $0\leqslant z\leqslant 1.75$ and the BAO
measurement at the largest redshift H(z = 2.34) taken after
\citep{Delubac} on the basis of BAO's in the Ly$\alpha$ forest of BOSS DR11 quasars. We define the following $\chi^2$ function, to be
minimized:
\begin{small}
\begin{equation}\label{xi2Omh2}
\chi^2_{Omh^2}=\sum_{i=1}^{N-1}\sum_{j=i+1}^{N}\left[\frac{Omh^2_{th}(H_i,H_j)-Omh^2_{obs}(H_i,H_j)}{\sigma_{Omh^2\,i,j}}\right]^2\,,
\end{equation}
\end{small}
where $N$ is the number of points $H(z)$ contained in the data set,
$H_i\equiv H(z_i)$, and the two-point diagnostic
$Omh^2(z_2,z_1)\equiv[h^2(z_2)-h^2(z_1)]/[(1+z_2)^3-(1+z_1)^3]$ was defined in\,\citep{SahShaSta},
with $h(z)=h\,E(z)$, and $\sigma_{Omh^2\,i,j}$ is the
uncertainty associated to the observed value $Omh^2_{obs}(H_i,H_j)$ for
a given pair of points $ij$, viz.\\
\begin{equation}
\sigma^2_{Omh^2\,i,j}=\frac{4\left[h^2(z_i)\sigma^2_{h(z_i)}+h^2(z_j)\sigma^2_{h(z_j)}\right]}{\left[(1+z_i)^3-(1+z_j)^3\right]^2}\,.
\end{equation}
For the  $\CC$CDM the two-point diagnostic boils down to $Omh^2(z_2,z_1)=\Om\,h^2$, which is constant for any pair
$z_1$, $z_2$. Using this testing tool and the
known observational information on $H(z)$ at the three redshift
values $z=0,0.57,2.34$  the aforementioned authors observed
that the average result is: $Omh^2=0.122\pm0.010$, with very little
variation from any pair of points taken. The obtained result is
significantly smaller than the  corresponding  Planck value of the
two-point diagnostic, which is constant and given by
$Omh^2=\Omo\,h^2=0.1415\pm0.0019$ \citep{Planck}.

A departure of $Omh^2$ from the Planck result should, according to
\citep{SahShaSta}, signal that the DE cannot be described by a rigid
cosmological constant.
For the $\CC$CDM we obtain $Omh^2=0.1250\pm0.0039$, and
$Omh^2=0.1402\pm0.0059$, by taking all data points, and excluding the
high redshift one, respectively.  Since there is a priori no reason to exclude the high-redshift point\,\citep{Delubac}, whose
uncertainty is one of the lowest in the full data sample, relaxing the tension with data may
require the dynamical nature of the DE. The vacuum models G1 and G2 considered here, Eqs.\,(\ref{eq:G1},\ref{eq:G2}), aim at cooperating in this task.

\begin{figure}[!t]
\includegraphics[scale=0.35]{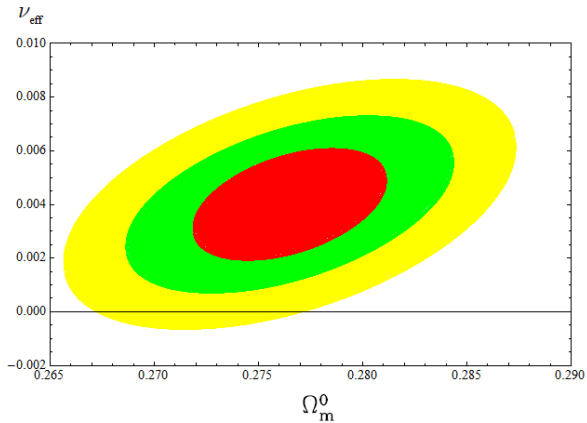}
\caption{\scriptsize Likelihood contours in the $(\Omega_m,\nueff)$ plane (for the values $-2\ln\mathcal{L}/\mathcal{L}_{max}=2.30$, $6.16, 11.81$, corresponding to 1$\sigma$, 2$\sigma$ and $3\sigma$ confidence levels for the G2 model using the full data analysis indicated in Table 2. The $\nueff=0$ region ($\CC$CDM) is disfavored at $\sim 3\sigma$.\label{fig:CL}}
\end{figure}

For these models  the theoretical value  $Omh_{th}^2$ of the two-point
diagnostic entering (\ref{xi2Omh2}) can be computed, in the matter-dominated epoch (relevant for such observable), as follows:
\begin{equation}\label{eq:Omh2zA}
Omh_G^2(z_i,z_j)=\frac{\Omo\,h^2}{\xi}\,\frac{\left(1+z_i\right)^{3\xi}-\left(1+z_j\right)^{3\xi}}{(1+z_i)^3-(1+z_j)^3}\,.
\end{equation}
It is evident that for $\xi=1$ we recover the $\CC$CDM
result, which remains anchored at $Omh^2(z_i,z_j)=\Omo h^2\ (\forall
z_i,\forall z_j)$. However, when we allow some small vacuum dynamics
(meaning $\nu$ and/or $\alpha$ different from zero) we
obtain a small departure of $\xi$ from $1$ and therefore the DE
diagnostic $Omh^2$ deviates from $\Omo h^2$. In this case $Omh^2$
evolves with cosmic time (or redshift).

To the above Hubble parameter data we add the recent supernovae type Ia data (SNIa), the Cosmic Microwave Background (CMB) shift parameter, the Baryonic Acoustic Oscillations (BAO's), the growth rate for structure formation (see next section) and the BBN and CMB anisotropy bounds. Contour lines for $\nueff=1-\xi$ are shown in Fig. 2 for model G2 at fixed $\xi'=1$. The $\chi^2$ functions associated to SNIa distance modulus
$\mu(z)$, the BAO $A$-parameter and the CMB shift parameter can be
found in \citep{GomSolBas}. Therein, one can also find the
corresponding references of the data sets that we have used in the
present analysis.

\begin{figure*}
\centering
\includegraphics[angle=0,width=0.318\linewidth]{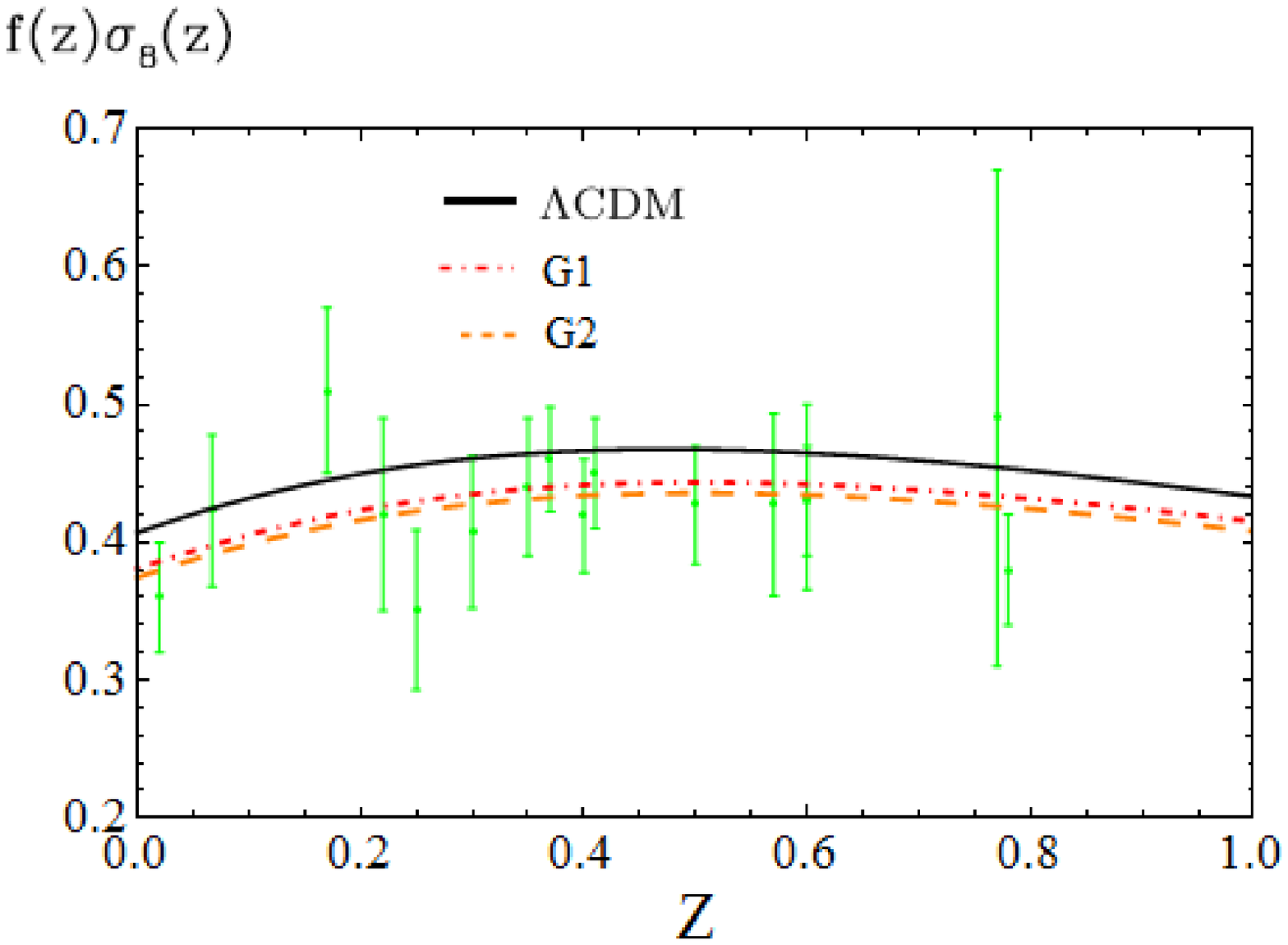}
\includegraphics[angle=0,,width=0.3\linewidth]{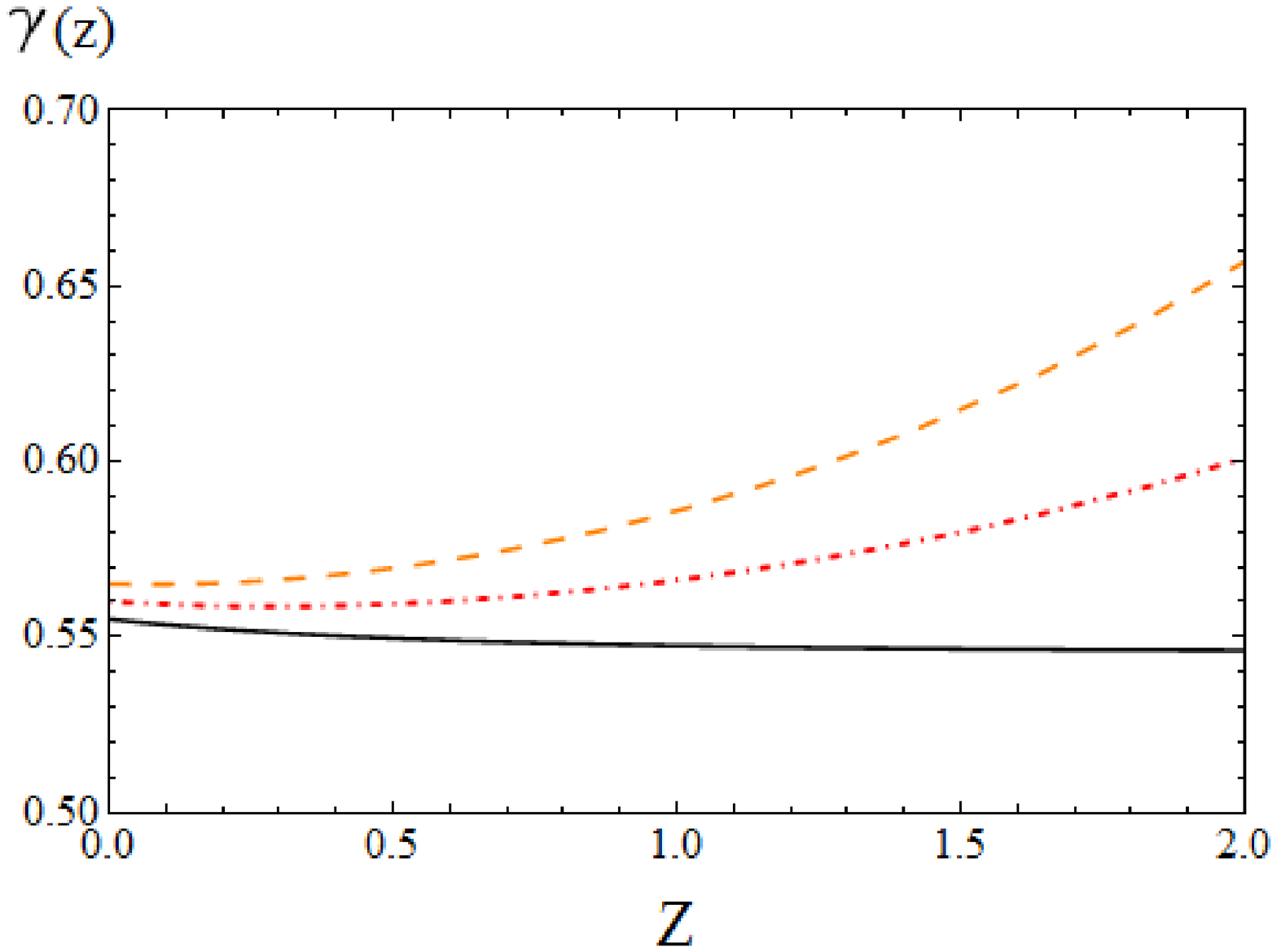}
\caption{\label{fig:LinearStructure}%
\scriptsize Left: Comparison of the observed data ֠with error bars (in green) ֠and the theoretical evolution of the weighted growth rate
of clustering $f(z)\sigma_8(z)$ for each dynamical vacuum model and the $\CC$CDM. Right: The corresponding evolution of the linear growth index $\gamma(z)$.
\vspace{0.3cm}
}
\end{figure*}


\section{Linear structure formation}\label{sect:LinStructure}

Finally, we take into consideration the data on the linear structure formation. For the G1 and G2 models the calculation of $\delta_m=\delta\rho_m/\rho_m$ is significantly  more
complicated than in the $\CC$CDM case and follows from
applying linear perturbation theory to Einstein's field equations
and Bianchi identity \eqref{eq:Bianchi}\,\citep{Grande2010}. The final result reads\footnote{The third-order feature of this equation is characteristic of the coupled systems of matter and DE perturbations for cosmologies with matter conservation, after eliminating the perturbations in the DE in favor of a single higher order equation for the matter part -- cf. \,\citep{self} for details. For $\CC=$const. Eq.\,\eqref{eq:thirdorder} boils down to the (derivative of the) second order one of the $\CC$CDM. }: \begin{equation}\label{eq:thirdorder}
\delta_m^{\prime\prime\prime}+\frac{\delta_m^{\prime\prime}}
{2a}(16-9\Omega_m)+\frac{3\delta_m^\prime}{2a^2}(8-11\Omega_m+3\Omega_m^2-a\Omega_m^\prime)=0\,,\end{equation}
with $\Omega_m(a)={8\pi\,G(a)}\rho_m(a)/{3H^2(a)}$. Notice that the $(\nu,\alpha)$  model-dependence is encoded in $H(a)$ -- cf. Eqs. (\ref{eq:DifEqH},\ref{eq:xixip}). To solve the above equation (numerically) we have to fix the initial
conditions for $\delta_m$, $\delta_m^\prime$ and
$\delta_m^{\prime\prime}$.
We  take due account of the fact that for these models at small $a$ (when non-relativistic matter
dominates over the vacuum) we have $\delta_m(a)=a^s$, where $s=3\xi-2=1-3\nueff$. If $\xi=1$ ($\nueff=0$), then $\delta_m(a)\sim a$
and we recover the $\CC$CDM behavior.
Thus, the initial conditions set at a high redshift
$z_i=(1-a_i)/a_i$, say $z_i=100$ (or at any higher value), are the
following. For the growth factor we have $\delta_m(a_i)=a_i^s$, and
for its derivatives: $\delta_m^\prime(a_i)=sa_i^{s-1}$,
$\delta_m^{\prime\prime}(a_i)=s(s-1)a_i^{s-2}$.

In practice we investigate the agreement with the structure formation data by comparing the theoretical linear growth prediction
%
$f(z)=-(1+z){d\ln\delta_m}/{dz}$
%
and the growth rate index $\gamma(z)$ with the available
growth data -- following \citep{GomSolBas} and references
therein. Recall that $\gamma$ is defined through $f(z)\simeq \Om(z)^{\gamma(z)}$, and one typically expects $\gamma(0)=0.56\pm 0.05$ for $\CC$CDM-like models\,\citep{Pouri2014}.
A most convenient related quantity is the weighted growth rate
$f(z)\sigma_8(z)$, cf.\,\citep{Percival08}, where
$\sigma_8(z)$ is the rms mass fluctuation amplitude on scales of
$R_8=8\,h^{-1}$ Mpc at redshift $z$. The latter is computed from
\begin{equation}\begin{small}\sigma_{\rm 8}(z)=\sigma_{8, \Lambda}
\frac{\delta_m(z)}{\delta^{\CC}_{m}(0)}
\left[\frac{\int_{0}^{\infty} k^{n_s+2} T^{2}(\Omega_{m},k)
W^2(kR_{8}) dk} {\int_{0}^{\infty} k^{n_s+2} T^{2}(\Omega_{m,
\Lambda},k) W^2(kR_{8}) dk}
\right]^{1/2}\label{s88general}\end{small}\end{equation}
$W$ being a top-hat smoothing function (see e.g. \citep{GomSolBas} for
details) and $T(\Omega_m,k)$ the transfer function, which we take from\,\citep{Bardeen}. The values of $\sigma_8\equiv\sigma_8(0)$ for the various models are collected in Table 1, and in Fig. 3 we plot $f(z)\sigma_8(z)$ and $\gamma(z)$ for them.

The joint likelihood analysis is performed on the set of Omh$^2$+BAO+SNIa+CMB, BBN and linear growth data,
involving one ($\Omega_m$) or two $(\Omega_m,\nueff)$ independently adjusted parameters depending on the model. For the $\CC$CDM we have one parameter ($n_p=1$) and for G1 and G2 we have $n_p=2$. Recall that for G2 we have fixed $\xi'=1$.

\section{Discussion}
The main results of this work are synthesized in Tables 1-2 and Figures 1-3. In particular, from Fig. 2 we see that the model parameter $\nueff$ for G2 is clearly projected onto the positive region, which encompasses most of the $3\sigma$ range. Remarkably, the $\chi^2$-value of the overall fit is smaller than that of $\CC$CDM for both G1 and G2 (cf. Tables 1-2).
To better assess the distinctive quality of the fits we apply the well known Akaike Information Criterion (AIC)\citep{Akaike}, which requires the condition $N_{\rm tot}/n_p > 40$ (amply satisfied in our case). It is defined, for Gaussian errors, as follows:
${\rm AIC}=-2\ln{\cal L_{\rm max}}+2n_{p}=\chi^2_{\rm min}+2n_{p}$, where ${\cal L_{\rm max}}$ (resp. $\chi^2_{\rm min}$) is the maximum (resp. minimum) of the likelihood (resp. $\chi^2$) function.
To test the effectiveness of models $M_i$ and $M_j$, one considers the pairwise difference
$(\Delta$AIC$)_{ij} = ({\rm AIC})_{i} - ({\rm AIC})_{j}$. The larger the
value of $\Delta_{ij}\equiv|\Delta({\rm AIC})_{ij}|$, the higher the evidence against the
model with larger value of ${\rm AIC}$, with $\Delta_{ij} \ge 2$ indicating a positive such evidence and $\Delta_{ij}\ge 6$
denoting significant such evidence.

From Tables 1-2 we see that when we compare the fit quality of models $i=$G1, G2 with that of  $j=\CC$CDM, in a situation when we take all the data for the fit optimization, we find $\overline{\Delta}_{ij}\equiv\Delta\overline{\rm AIC}\gtrsim 9.43$ for G2 and $4.72$ for G1, suggesting significant evidence in favor of these models (especially G2) against the $\CC$CDM -- the evidence ratio\,\citep{Akaike} being ${\rm ER}=e^{\overline{\Delta}_{ij}/2}\gtrsim111.6$ for G2 and $10.6$ for G1. Worth noticing is also the result of the fit when we exclude the growth data from the fitting procedure but still add their contribution to the total $\chi^2$. This fit is of course less optimized, but allows us to risk a prediction for the linear growth and hence to test the level of agreement with these data points (cf. Fig. 3). It turns out that the corresponding AIC pairwise difference with the $\CC$CDM are similar as before (cf. Tables 1-2). Therefore, the $\CC$CDM appears significantly disfavored versus the dynamical vacuum models, especially in front of G2, according to the Akaike Information Criterion. Let us mention that if we remove all of the $H(z)$ data points from our analysis the fit quality weakens, but it still gives a better fit than the $\CC$CDM (cf. the third and fourth row of Tables 1 and 2). If, however, we keep these data points but remove \textit{only} the high redshift point $z=2.34$\,\citep{Delubac}, the outcome is not dramatically different from the previous situation (confer the starred scenarios in Tables 1 and 2), as in both cases the significance of $\nueff\neq0$ is still close to $\sim 2\sigma$ with $\Delta_{ij}>7$ for G2 (hence still strongly favored, with ${\rm ER}> 33$). In this sense the high $z$  point may not be so crucial for claiming hints in favor of dynamical vacuum, as the hints themselves seem to emerge more as an overall effect of the data. While we are awaiting for new measurements of the Hubble parameter at high redshift to better assess their real impact, we have checked that if we add to our analysis the points $z=2.30$\,\citep{Busca} and  $z=2.36$\,\citep{FontRibera}, not included in either \citep{SahShaSta} or \citep{Ding}, our conclusions remain unchanged. Ditto if using the three high $z$ points only.


To summarize, our study singles out a general class of vacuum models, whose dynamical behavior challenges the overall fit quality of the rigid $\CC$-term inherent to the concordance $\CC$CDM model. From the data on expansion, structure formation, BBN and CMB observables we conclude that the $\CC$CDM model is currently disfavored at $\sim 3\sigma$ level as compared to the best dynamical ones.

\section{Acknowledgements}

We thank the anonymous referee for his/her thorough report on our work and for very useful suggestions to improve our analysis. JS has been supported 
in part by MICINN, CPAN and Generalitat de Catalunya; AGV acknowledges support by APIF grant of the U. Barcelona.

\newcommand{\CQG}[3]{{ Class. Quant. Grav. } {\bf #1} (#2) {#3}}
\newcommand{\JCAP}[3]{{ JCAP} {\bf#1} (#2)  {#3}}
\newcommand{\APJ}[3]{{ Astrophys. J. } {\bf #1} (#2)  {#3}}
\newcommand{\AMJ}[3]{{ Astronom. J. } {\bf #1} (#2)  {#3}}
\newcommand{\APP}[3]{{ Astropart. Phys. } {\bf #1} (#2)  {#3}}
\newcommand{\AAP}[3]{{ Astron. Astrophys. } {\bf #1} (#2)  {#3}}
\newcommand{\MNRAS}[3]{{ Mon. Not. Roy. Astron. Soc.} {\bf #1} (#2)  {#3}}
\newcommand{\PR}[3]{{ Phys. Rep. } {\bf #1} (#2)  {#3}}
\newcommand{\RMP}[3]{{ Rev. Mod. Phys. } {\bf #1} (#2)  {#3}}
\newcommand{\JPA}[3]{{ J. Phys. A: Math. Theor.} {\bf #1} (#2)  {#3}}
\newcommand{\ProgS}[3]{{ Prog. Theor. Phys. Supp.} {\bf #1} (#2)  {#3}}
\newcommand{\APJS}[3]{{ Astrophys. J. Supl.} {\bf #1} (#2)  {#3}}

\newcommand{\Prog}[3]{{ Prog. Theor. Phys.} {\bf #1}  (#2) {#3}}
\newcommand{\IJMPA}[3]{{ Int. J. of Mod. Phys. A} {\bf #1}  {(#2)} {#3}}
\newcommand{\IJMPD}[3]{{ Int. J. of Mod. Phys. D} {\bf #1}  {(#2)} {#3}}
\newcommand{\GRG}[3]{{ Gen. Rel. Grav.} {\bf #1}  {(#2)} {#3}}



\begin{thebibliography}{}

\bibitem[Planck Collaboration XIII (2015)]{Planck}
Ade, P.A.R, et al. [Planck Collaboration], \textit{Planck 2015
results. XIII. Cosmological parameters}, arXiv:1502.01589.

\bibitem[Akaike (1974)]{Akaike}
Akaike, H. IEEE Transactions of Automatic Control,
  { 19} (1974) 716;
  N. Sugiura, Communications in Statistics A, Theory and Methods, {7} (1978) 13; K.P. Burnham \& D.R. Anderson, \textit{Model selection and multimodel
inference} (Springer, New York, 2002).

\bibitem[Bardeen et al. (1986)]{Bardeen}
Bardeen, J. M.,  Bond, J. R.,  Kaiser, N., \& Szalay,  A. S.,
ApJ {\bf 304} (1986) 15.

\bibitem[Basilakos \& Sol\`{a} (2014)]{Mirage} Basilakos, S.,  \& Sol\`a, J., Mon. Not. Roy. Astron. Soc. {\bf 437} (2014) 3331

\bibitem[Busca et al. (2013)]{Busca}
Busca, N.G. et al.,  
A\&A {\bf 552} (2013) A96.

\bibitem[Chiba (2011)]{Chiba} 
Chiba, T., Prog. Theor. Phys. {\bf 126} (2011) 993;
Nagata, R., Chiba, T., \& Sugiyama, N.,  Phys. Rev. D{\bf 69} (2004) 083512.

\bibitem[Delubac (2015)]{Delubac}
Delubac, T., et al., A\&A  {\bf 574}
(2015) A59.

\bibitem[Ding {et~al.}(2015)]{Ding}
Ding, X., et al., ApJ 803 (2015)  L22.



	
\bibitem[Farooq \& Ratra (2013)]{Farooq}
Farooq, O., \& Ratra, B., ApJ {\bf 766} (2013) L7.

\bibitem[Font-Ribera et al. (2014)]{FontRibera}
Font-Ribera, A. et al.,  JCAP {\bf 05} (2014) 027.


\bibitem[G\'omez-Valent, Sol\`{a} \& Basilakos (2015)]{GomSolBas}
G\'omez-Valent, A., Sol\`{a}, J., \& Basilakos, S., JCAP {\bf 01} (2015) 004; A. G\'omez-Valent \&  J. Sol\`{a}, MNRAS  {\bf 448} (2015) 2810.

 \bibitem[G\'omez-Valent, Karimkhani \& Sol\`a (2015)]{self} G\'omez-Valent, A.,  Karimkhani, E.,  \&  Sol\`{a}, J.,
 work in preparation.

\bibitem[Grande {et~al.}(2010)]{Grande2010}	
Grande, J., et. al, Class. Quant. Grav. {\bf 27} (2010) 105004;
JCAP 08  (2011) 007.

\bibitem[Lima, Basilakos \& Sol\`a (2013)]{LimBasSol}
Lima, J. A. S., Basilakos, S., \& Sol\`a, J., MNRAS {\bf 431} (2013)  923; Gen. Rel. Grav. {\bf 47} (2015) 40.

\bibitem[Padmanabhan (2003)]{CCP2} Padmanabhan, T., Phys. Rept. {\bf 380} (2003) {235}.

\bibitem[Peebles \& Ratra (2003)]{CCP3}   Peebles P.J.E., \&
Ratra, B., Rev. Mod. Phys. {\bf 75} (2003) 559.

\bibitem[Perlmutter {et~al.}(1999)]{SNIaPerl}
Perlmutter, S., et al.
  ApJ.  {\bf 517} (1999) 565.
	
\bibitem[Pouri, Basilakos \& Plionis (2014)]{Pouri2014} Pouri, A., Basilakos, S., \& Plionis, M.,  JCAP {\bf 08} (2014) 042.

\bibitem[Riess {et~al.}(1998)]{SNIaRiess}
Riess, A.~G., ~et al. ,
  Astron.\ J.\  {\bf 116} (1998) 1009.

\bibitem[Sahni, Shafieloo \& Starobinsky (2014)]{SahShaSta}
Sahni, V., Shafieloo, A., \& Starobinsky, A. A., ApJ. {\bf 793}  (2014) L40.
	
\bibitem[Sahni \& Starobinsky (2000)]{CCP1}
Sahni, V., \& Starobinsky, A. A., Int. J. of Mod. Phys. A{\bf 9} {(2000)} {373}.

\bibitem[Shafieloo {et~al.} (2006)]{Shafieloo2006}
Shafieloo, A., Alam, U., Sahni, V., Starobinsky, A. A., Mon. Not. Roy. Astron. Soc. {\bf 366} (2006) 1081.

\bibitem[Shapiro \& Sol\`a (2009)]{ShapSol09}
Shapiro, I. L., \& Sol\`a, J., Phys.Lett. B{\bf 682} (2009) 105;
JHEP 02 (2002) 006.

\bibitem[Sol\`a (2008)]{Fossil07} Sol\`a, J.,
{J. of Phys.}  A{\bf 41} (2008) {164066}.

 	

\bibitem[Sol\`a (2013)]{JSPRev2013}
Sol\`a, J., 
J. Phys. Conf. Ser. {\bf 453} (2013) 012015.

\bibitem[Sol\`a \& G\'omez (2015)]{SolGom2015} Sol\`{a}, J., \& G\'omez-Valent, A., Int.J.Mod.Phys. D{\bf 24} (2015) 1541003.

\bibitem[Sol\`a (2015)]{Sola2015}
Sol\`a, J., 
arXiv:1505.05863 (to appear in Int. J. of Mod. Phys. D).

\bibitem[Sol\`a \& Stefancic (2006, 2005)]{SolaStefancic} Sol\`a, J., \& Stefancic, H.,  Mod. Phys. Lett. A{\bf 21} (2006) 479; Phys.Lett. B{\bf 624} (2005) 147.

\bibitem[Song \& Percival (2009)]{Percival08} Song, Y-S., \& Percival, W.J., JCAP {\bf 10} (2009) 004.

\bibitem[Uzan (2011)] {Uzan} 	
Uzan, J-P.,  Living Rev.Rel. {\bf 14} (2011) 2.

\bibitem[Weinberg (1989)]{Weinberg}
 Weinberg, S., Rev. Mod. Phys. {\bf 61} {(1989)} {1}.







\end{thebibliography}
\end{document}